Comments on the :

Observation of Long-Range Near-Side Angular Correlations in proton-proton Collisions at the LHC"
CMS Collaboration.
ArXiv:1009.4122[hep-ex]

L. J. Gutay, R. P. Scharenberg and B. K. Srivastava
Department of physics
Purdue University
West Lafayette, IN-47907,USA

The paper presented the results for the observation of long range correlations (LRC) in p-p collisions at LHC energies. It is mentioned that this is the first observation of , near side feature in two particle correlation functions in p-p or p-pbar collisions. The paper concludes that the physical origin is not yet understood.

UA5 experiment at CERN has observed the LRC in pbar-p collisions at 200, 546 and 900 GeV. The correlations were understood in terms of a cluster model [1].
The E-735 experiment at Fermi lab also studied the LRC in p-pbar collisions at 1.8TeV [2]. In both experiments UA5 and E735 the absolute magnitude of the correlation strength was determined. It was shown that magnitude of the LRC increases with increase in the collision energy. E735 has also interpreted the results in terms of a cluster model of the production. It was also shown that the variance of the asymmetry , which gives the cluster size, increases with the increase in collision energy. The variance was also found to depend on the multiplicity. In the later publication it was shown that the deconfined phase was created in pbar-p at 1.8 TeV [3].

Recently, STAR has published the measurement of LRC in Au+Au collisions at 200 GeV/A[4]. It has been shown that LRC are due to the presence of multi-parton interactions . The origin of these long range correlations have been understood. Two phenomenological models the Dual Parton Model (DPM) and Color Glass Condensate (CGC) had already predicted the existence of LRC [5,6].